\pdfoutput=1

\documentclass[11pt]{article}

\usepackage[preprint]{acl}

\usepackage{times}
\usepackage{latexsym}

\usepackage[T1]{fontenc}

\usepackage[utf8]{inputenc}

\usepackage{microtype}

\usepackage{inconsolata}

\usepackage{graphicx}
\usepackage{listings}

\usepackage{booktabs}
\usepackage[table,xcdraw]{xcolor}
\definecolor{codegreen}{rgb}{0,0.6,0}
\definecolor{codegray}{rgb}{0.5,0.5,0.5}
\definecolor{codepurple}{rgb}{0.58,0,0.82}
\definecolor{backcolour}{rgb}{0.95,0.95,0.92}

\title{Repairing Tool Calls Using Post-tool Execution Reflection and RAG}

\author{
 \textbf{Jason Tsay},
 \textbf{Zidane Wright},
 \textbf{Gaodan Fang},
 \textbf{Kiran Kate},
\\
 \textbf{Saurabh Jha},
 \textbf{Yara Rizk}
\\
\\
 IBM Research
\\
 \small{
   \textbf{Correspondence:} \href{mailto:jason.tsay@ibm.com}{jason.tsay@ibm.com}
 }
}

\begin{document}
\maketitle
\begin{abstract}

Agentic systems interact with external systems by calling tools such as Python functions, REST API endpoints, or command line tools such as \emph{kubectl} in Kubernetes. These tool calls often fail for various syntactic and semantic reasons. Some less obvious semantic errors can only be identified and resolved after analyzing the tool's response. To repair these errors, we develop a post-tool execution reflection component that combines large language model (LLM)-based reflection with domain-specific retrieval-augmented generation (RAG) using documents describing both the specific tool being called and troubleshooting documents related to the tool. For this paper, we focus on the use case of the \emph{kubectl} command line tool to manage Kubernetes, a platform for orchestrating cluster applications. Through a larger empirical study and a smaller manual evaluation, we find that our RAG-based reflection will repair kubectl commands such that they are both more likely to successfully execute (\textit{pass rate}) for 55\% of our models evaluated and 36\% more likely to correctly answer the user query on average. We find that troubleshooting documents improve \textit{pass rate} compared to official documentation by an average of 10\%.

\end{abstract}

\section{Introduction}

When large language models (LLMs) call tools to address user queries, they identify the tool name and assign values to input arguments \cite{qin2023toolllm, patil2023gorilla}. Whether the tool is a REST API endpoint, Python function, or command line tool, they format the call accordingly and send it to the appropriate tool executor.

These tool calls sometimes fail for various reasons. For example, the LLM may have hallucinated the function name or made syntactical errors in formulating the tool call \cite{deshpande2025trail}. Such syntactic errors can be identified before the tool executor attempts to execute the call. However, 
some errors are not as obvious such as semantic or contextual errors. Such errors can only be identified and resolved after executing the tool and analyzing the tool's response \cite{sun2024toolsfaildetectingsilent}. An example contextual error is when attempting to retrieve a list of pods in a given namespace in Kubernetes, it is only possible to know that the namespace doesn't exist after executing that command. 

In agentic settings, post-tool execution errors are usually handled by adding the tool response to the prompt in the next turn and allowing agents to attempt a resolution based on their reasoning abilities \cite{aksitov2023rest, yao2023reactsynergizingreasoningacting}. Often, the LLMs are prompted to reflect and reason explicitly on these outcomes and change its course of action, we call this self-reflection. Even though different models have exhibited varying levels of reasoning sophistication \cite{cheng2025empowering}, most agents still fail at recovering from tool errors \cite{basu2025nestful}. This suggests that LLMs underlying knowledge of tools (if any) may be insufficient for real-time troubleshooting. Thus, we believe components beyond self-reflection that integrate domain knowledge are necessary for agents to identify and repair errors when they happen.

\begin{figure*}[!tbh]
  \centering
  \includegraphics[width=0.85\textwidth]{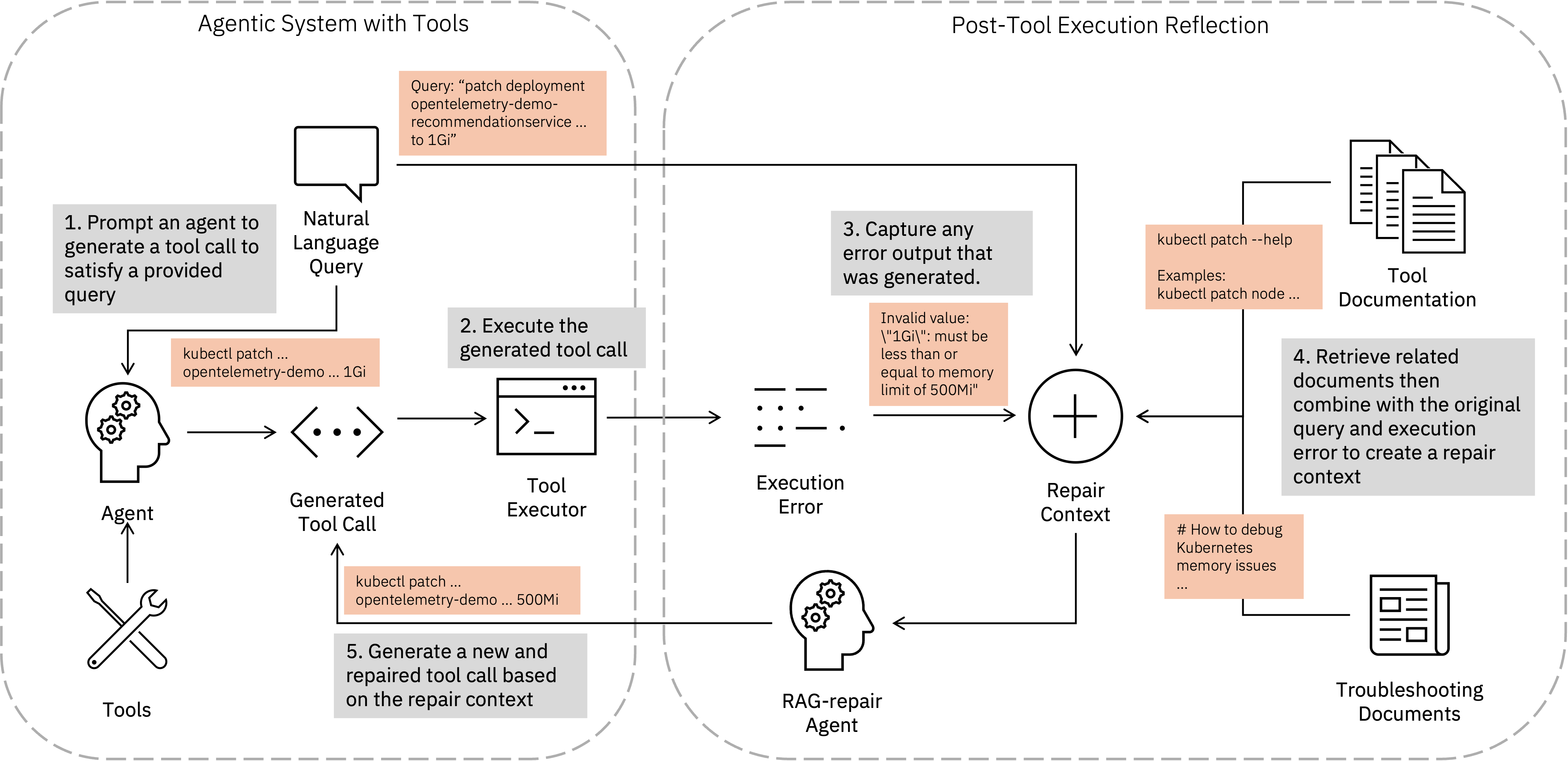}
  \caption{The process of RAG-based repair. On the left we have an agentic system that generates tool calls and executes them using some tool executor. The right side shows our RAG-Repair approach for post-tool execution reflection that uses the repair context to repair the failing tool cool. }
  \label{fig:overview}
\end{figure*}

Our approach 
combines contextual information from the environment through the results of the tool execution along with additional domain information from external documents using RAG (Retrieval Augmented Generation). Figure~\ref{fig:overview} shows the workflow of such an agentic system. The agent first generates a tool call based on a user's natural language query. This tool call is then executed, and any resulting execution errors are captured. When an error occurs, the system starts to construct a ``repair context'' using a RAG-based process. Using the natural language query, the captured error output and the generated tool call as a key, we retrieve relevant information from official tool documentation, troubleshooting documents, and any other potentially useful information for recovering from the error. We combine that key and the retrieved results to create the repair context and pass it to a repair agent to generate a repaired tool call. This repaired tool call can be re-executed and the RAG-based repair process may be repeated as necessary.

To evaluate our post-tool reflection using RAG, we use a dataset of natural language to Kubernetes commands (called \emph{kubectl}). 
This is a real-world tool usage dataset with complex tools which can be executed to obtain execution errors. The \emph{kubectl} commands also have the benefit of being relatively bounded and executable in ways that may or may not change the state of the external environment. We evaluate our reflection component using a set of 772 failing \emph{kubectl} commands which are generated from natural language user queries by a baseline agent using an LLM. We perform both a larger-scale empirical evaluation on how often our reflection is able to repair commands such that they are able to successfully execute (\textit{pass rate}) and a smaller-scale manual evaluation where we examine how correctly the repaired command fulfills the user query. Our findings suggest that compared to our non-RAG baseline, both \textit{pass rate} and correctness are more likely to improve using our RAG-based reflection. To examine the effect of document choice for the RAG component, we also perform an ablation study where we find that troubleshooting documents tend to increase \textit{pass rate} compared to official documentation.

An open source implementation of this approach is available as part of the Agent Lifecycle Toolkit\footnote{https://github.com/AgentToolkit/agent-lifecycle-toolkit}.

\section{Related Work}
\subsection{Tool Calling Errors}
Most of the tool calling work focuses on the ability of LLMs to generate a correct tool call given a catalog of tool specifications~\citep{qin2023toolllm, patil2023gorilla, abdelaziz2024granite}. Studies on tool calling errors~\citep{kokane2024spectoolbenchmarkcharacterizingerrors, kokane2025toolscan, winstontaxonomy} emphasize categories related to generation of tool calls. 
Post-tool execution errors can be categorized into silent errors~\citep{sun2024toolsfaildetectingsilent} and explicit execution failures~\citep{kokane2025toolscan}. A prevalent approach in handling execution failures is to pass the error to the LLM/agent and rely on its ability to correct the course of action~\citep{yao2023reactsynergizingreasoningacting, qin2023toolllm}. Unlike this approach, our work leverages the knowledge base of tool documentation and troubleshooting discussions to help the LLM recover from execution failures.

\subsection{RAG + code documentation/error logs}
While most popular for text tasks, RAG has been effective in code generation~\citep{wang-etal-2025-coderag} and performance bug fixes~\citep{garg2025rapgenapproachfixingcode}. For tool calling, RAG has been used to select a subset of tool specifications from a large pool in order to add those to the LLM prompt~\citep{qin2023toolllm, lumer2024toolshed}. To the best of our knowledge, no existing work studies RAG for tool execution errors.


\section{RAG for Post-Tool Execution Reflection}

Even though tool augmented LLMs are given the tool specifications in the prompts, those are often limited to tool descriptions, parameter names, descriptions, and types. Providing more information about all tools in the catalog is not practical considering the models' limited effective context window.
Therefore, LLMs and LLM-based agents may lack specific troubleshooting knowledge for a given tool call and error, specially for tools they have not seen in their training data. 
RAG affords the ability to provide a large collection of domain-specific documents and knowledge and the ability 
to extract only the relevant or noteworthy parts to the LLM prompt to fix a tool execution error. Such RAG-based post tool execution repair (hereafter RAG-Repair) is a promising approach to handle such errors.



In this paper, we create the following RAG-Repair pipeline (see Figure~\ref{fig:overview} for an overview). We assume that when an agent or LLM calls an external tool, there is an associated natural language query, usually from the user. We also assume that the tool call is comprised of a target and a set of required and/or optional parameters. Finally, we assume that a response from the external tool is available and contains some natural language describing the error (example: ``Error: Invalid Value''). Then we use a combination of LLM-based reflection and contextual information from the tool response along with domain-specific information from RAG to generate a repaired tool call for re-execution. 

For RAG-Repair, we currently focus on two types of documents to provide domain knowledge: 1) official documentation of the tools and 2) troubleshooting documents. For evaluation, we consider the task of natural language query to \emph{kubectl} command-line tool calls for Kubernetes. We use the official Kubernetes documentation\footnote{https://kubernetes.io/docs/reference/kubectl/ \label{k8s-docs}} and scrape 101 manual pages, each describing a specific \emph{kubectl} command or more specific versions of commands (e.g., \emph{kubectl create} vs \emph{kubectl create service}) along with possible arguments. 
For troubleshooting documents, we focus on \emph{kubectl}-related StackExchange questions from the RedPajama dataset~\cite{together2023redpajama}. Questions were filtered by a threshold question score of 5 and mention of any valid \emph{kubectl [command]} from the documentation\footref{k8s-docs} for a final set of 186 troubleshooting documents.
We use the ChromaDB vector database and MPNet~\cite{mpnet} for sentence transformation with a chunk size of 1,500 and default distance metric of Squared L2.

\section{Experiments and Results}


\subsection{Evaluation Dataset}

For this paper, we focus on the domain of Kubernetes commands because it is a popular yet relatively bounded domain where execution data, robust official documentation, and troubleshooting documents are readily available. 
We use the SRE subset from ITBench~\cite{jha2025itbench} (Apache2.0) to provide a set of Kubernetes-related natural language user queries and their baseline SRE-Agent to generate \emph{kubectl} commands from the queries. 
We then only consider commands that result in execution failures and involve the OpenTelemetry Demo (otel-demo) that we use in our environment. 
The final evaluation dataset contains 772 sets of queries, generated commands, and execution errors. 

\subsection{Evaluation Methodology}

\begin{table}[!tb]
\centering
\footnotesize
\begin{tabular}{@{}lllc@{}}
\toprule
\textbf{Model}          & \textbf{Baseline} & \textbf{RAG} & \textbf{$\Delta$}                \\ \midrule
\textbf{Llama3.1 8b}    & 0.557              & 0.627                & \cellcolor[HTML]{9AFF99}0.070  \\
\textbf{Granite3.2 8b}  & 0.557              & 0.387                & \cellcolor[HTML]{FFCCC9}-0.170 \\
\textbf{Mixtral 8x7b}   & 0.196              & 0.344                & \cellcolor[HTML]{9AFF99}0.148  \\
\textbf{Phi4}           & 0.329              & 0.349                & \cellcolor[HTML]{FFFFFF}0.017  \\
\textbf{Llama4Maverick} & 0.404              & 0.497                & \cellcolor[HTML]{9AFF99}0.093 \\
\textbf{DeepSeek-V3}    & 0.524              & 0.324                & \cellcolor[HTML]{FFCCC9}-0.200 \\
\textbf{Mixtral 8x22b}  & 0.461              & 0.454                & -0.007                         \\
\textbf{Llama3.3 70b}   & 0.422              & 0.552                & \cellcolor[HTML]{9AFF99}0.130  \\
\textbf{Qwen2.5 72b}    & 0.397              & 0.403                & 0.006                          \\
\textbf{Mistral Large}  & 0.441              & 0.144                & \cellcolor[HTML]{FFCCC9}-0.297 \\
\textbf{GPT-4o}         & 0.488              & 0.451                & -0.037                         \\ \bottomrule
\end{tabular}
\caption{Repair \textit{pass rate} in ascending order of model size. Baseline is a non-RAG reflection compared to RAG-Repair which uses RAG and both official documentation and troubleshooting documents. Increases or decreases in \textit{pass rate} over 5\% are highlighted.}
\label{tab:largeresults}
\end{table}

\begin{table*}[!htb]
\scriptsize
\begin{tabular}{@{}lllllllllllllllll@{}}
                   & \multicolumn{2}{l}{\textbf{Llama3.1 8b}} & \multicolumn{2}{l}{\textbf{Llama3.3 70b}} & \multicolumn{2}{l}{\textbf{Llama4Mav}} & \multicolumn{2}{l}{\textbf{GPT4o}} & \multicolumn{2}{l}{\textbf{DeepSeekv3}} & \multicolumn{2}{l}{\textbf{Mixtral 8x22b}} & \multicolumn{2}{l}{\textbf{Granite3.2 8b}} & \multicolumn{2}{l}{\textbf{Average}} \\ 
                   & Base                & RAG                & Base                & RAG                 & Base               & RAG               & Base             & RAG             & Base                & RAG                & Base                 & RAG                 & Base                 & RAG                  & Base              & RAG              \\ \midrule
\textbf{Correct}   & 0.04                & \cellcolor[HTML]{9AFF99}0.28               & 0.36                & 0.32                & 0.44               & \cellcolor[HTML]{9AFF99}0.52              & 0.36             & \cellcolor[HTML]{9AFF99}0.44            & 0.40                & 0.40               & 0.28                 & 0.32                & 0.12                 & \cellcolor[HTML]{9AFF99}0.24                 & 0.29              & \cellcolor[HTML]{9AFF99}0.36             \\
\textbf{Partial}   & 0.04                & 0.08               & 0.04                & 0.04                & 0.08               & 0.04              & 0.16             & \cellcolor[HTML]{FFCCC9}0.04            & 0.04                & 0.08               & 0.00                 & 0.04                & 0.00                 & 0.00                 & 0.05              & 0.05             \\
\textbf{Adjacent}  & 0.12                & \cellcolor[HTML]{FFCCC9}0.00               & 0.08                & 0.12                & 0.08               & 0.04              & 0.04             & 0.00            & 0.04                & 0.04               & 0.12                 & \cellcolor[HTML]{FFCCC9}0.04                & 0.08                 & 0.08                 & 0.08              & 0.05             \\
\textbf{Incorrect} & 0.52                & \cellcolor[HTML]{9AFF99}0.28               & 0.16                & 0.20                & 0.16               & \cellcolor[HTML]{FFCCC9}0.28              & 0.00             & \cellcolor[HTML]{FFCCC9}0.12            & 0.40                & \cellcolor[HTML]{9AFF99}0.16               & 0.28                 & \cellcolor[HTML]{9AFF99}0.20                & 0.36                 & \cellcolor[HTML]{9AFF99}0.20                 & 0.27              & \cellcolor[HTML]{9AFF99}0.21             \\
\textbf{Error}     & 0.28                & \cellcolor[HTML]{FFCCC9}0.36               & 0.36                & 0.32                & 0.24               & \cellcolor[HTML]{9AFF99}0.12              & 0.44             & 0.40            & 0.12                & \cellcolor[HTML]{FFCCC9}0.32               & 0.32                 & \cellcolor[HTML]{FFCCC9}0.40                & 0.44                 & 0.48                 & 0.31              & 0.34             \\ 
\end{tabular}
\caption{Summary of manual correctness evaluation where rates for each correctness category are reported (see Appendix for details). Correctness category increases or decreases over 5\% are highlighted.}
\label{tab:manualresults}
\end{table*}

This dataset is used to perform a larger-scale empirical evaluation of how often RAG-Repair can take a failing kubectl command and repair the command such that it will successfully execute (\textit{pass rate}), specifically if the return code from execution is 0. 
We evaluate RAG-Repair on 11 LLMs of different sizes and families and also compare it with a self-reflection style baseline.

To understand what kind of documents help RAG-Repair, we perform an ablation study to compare RAG with only official \emph{kubectl} documentation to one that only uses troubleshooting documents. We select five LLMs which had the highest \textit{pass rates} from the earlier experiment.

Automatic evaluation using \textit{pass rate} has flaws as a command may execute without errors but may not correctly address the user's query. So we perform manual evaluation to check if the command satisfies the user requirement. For a sample of 25 unique errors, we manually annotated the repair outputs using the following five categories: 
1) \emph{Correct} - the user query was directly answered, 
2) \emph{Partially Correct} - only part of the user query was directly answered, 
3) \emph{Adjacent} - the result does not directly answer query but is a reasonable first step, 
4) \emph{Incorrect} - the execution succeeded but did not answer the user query in any way, and 
5) \emph{Error} - the execution failed. 
We validated this evaluation setup by having two authors independently annotate the same 25 cases and calculated inter-rater reliability with a weighted Cohen's kappa of 0.79. 



\subsection{Results}

Table~\ref{tab:largeresults} summarizes the \textit{pass rate} comparing a non-RAG baseline and RAG-Repair. Using RAG improves \textit{pass rate} for six of the eleven models. We note particularly poor performance for large models like Mistral Large and DeepSeek-V3. Manually examining DeepSeek-V3 (see Table~\ref{tab:manualresults}) suggests that it tends to hallucinate parameters, e.g., returning `kubectl expose pod hello-node-64c578bdf8-jp7dt' for a query to create a service called ``catalog-service-clusterip.'' 

Table~\ref{tab:ablationresults} summarizes the RAG ablation study. We note that for three out of the five models, a RAG component that only contains troubleshooting documents (StackExchange posts related to \emph{kubectl}) have the highest performance, suggesting that official manual pages are actually reducing \textit{pass rate} (which is true in three out of five models when comparing non-RAG to manual-only). 

\begin{table}[!th]
\centering
\footnotesize
\begin{tabular}{@{}lllll@{}}
\toprule
\textbf{Model} & \textbf{None} & \textbf{Man} & \textbf{TS} & \textbf{Full} \\
\midrule
\textbf{Llama3.1 8b}    & 0.56           & 0.63             & \textbf{0.64}   & 0.63             \\
\textbf{Llama4Maverick} & 0.40           & 0.43             & 0.49            & \textbf{0.50}    \\
\textbf{Llama3.3 70b}   & 0.42           & 0.34             & \textbf{0.58}   & 0.55             \\
\textbf{Mixtral 8x22b}  & 0.46           & 0.39             & \textbf{0.47}   & 0.45             \\
\textbf{GPT-4o}         & \textbf{0.49}  & 0.37             & 0.47            & 0.45             \\
\bottomrule
\end{tabular}
\caption{Ablation study for RAG documents, reporting \textit{pass rate} for four conditions: \emph{None} (baseline without RAG-based repair), \emph{Man} (contains official kubectl \emph{man}ual pages), \emph{TS} (contains TroubleShooting documents), and \emph{Full}. Bolded \textit{pass rate} is the best performing for a given model.}
\label{tab:ablationresults}
\end{table}

Table~\ref{tab:manualresults} summarizes the small-scale manual evaluation. We note that when comparing the baseline to RAG-Repair, five out of seven models see improvement in the ``Correct'' category, on average 7\% with the largest improvement with our two smallest models: Llama3.1 8b at 24\% and Granite3.2 8b at 12\%. This suggests that small models may be missing kubectl-related data in training. An example is that a query requests alerts in a namespace and GPT-4o generates ``kubectl get alerts'' which hallucinates the resource ``alerts.'' For repair using Llama3.1, not providing documents will generate the syntactically valid but incorrect command ``kubectl api-versions'' while providing the manual page for the correct command ``kubectl get events'' will correctly repair the command. Conversely, a large model like GPT-4o is able to correct this command with or without documents, suggesting it already contains this knowledge but reflection was required to make use of it. 
Overall, this evaluation suggests that even for models where \textit{pass rate} does not improve such as Granite and Mixtral, correctness still improves (or at least does not decrease in the case of DeepSeek-V3).

\section{Conclusion}

This paper presents a RAG-Repair post-tool execution reflection component that uses domain knowledge to assist in repairing failing \emph{kubectl} tool calls to improve on both execution success in terms of \textit{pass rate} and correctness in answering user queries. Although our chosen use case and domain of Kubernetes is promising, future work will examine this approach for other tool-calling domains such as particular REST APIs or Python functions. The ablation study also suggests further study of the types and usage of documents, particularly official documentation. We expect repair performance to improve as we better understand the relationship between tool reflection and domain knowledge.

\clearpage
\section*{Limitations}

A significant limitation of this paper is that we solely focus on the use case of \emph{kubectl} and Kubernetes. This limits our ability to claim how generalizable this reflection approach is. This is especially the case as our approach requires the inclusion of domain-specific documents and knowledge (in this case official documentation and StackExchange questions). Although collecting similar documents for other domains is reasonable, we have not tested doing so for this paper. 
Another limitation is that evaluating ``correctness'' particularly in this domain is difficult. For many user queries, there are multiple equally valid approaches to answer them. An example of this from our small-scale dataset is searching for ``opentelemetry-demo-grafana'' across all deployments (which is not normally allowed in \emph{kubectl}). One way to do this is to use the field selector parameter to filter by deployment name (``kubectl get deployment -A --field-selector=metadata.name=opentelemetry-demo-grafana'') but another approach (used by RAG-Repair) was to use the \emph{grep} tool in conjunction with \emph{kubectl} (``kubectl get deployments --all-namespaces | grep opentelemetry-demo-grafana''). In our case, we decided to mitigate this difficulty using a manual evaluation but other, more scalable methods of evaluation may be possible.
Both our larger-scale \textit{pass rate} evaluation and manual annotation have their limitations. \textit{Pass rate} simply looks if the command successfully executes with a return code of 0, not the semantic correctness. Manual annotation is inherently not scalable and involves subjective judgments and domain knowledge from the annotator. We attempt to mitigate this limitation by having two annotators and calculating inter-rater reliability. Including a small-scale manual evaluation is our attempt to mitigate the limitation of the \textit{pass rate} metric by examining correctness along with execution success. 
One possible risk of this work is that a well-performing tool call repair component may enable malicious usage of external tools. One simple example may be an agent that performs DDoS attacks on an external service. Our approach could potentially help this agent to mutate and repair malicious tool calls on the fly. 

\bibliography{citations}

\appendix
\lstdefinestyle{listingstyle}{
    backgroundcolor=\color{backcolour},   
    commentstyle=\color{codegreen},
    keywordstyle=\color{magenta},
    numberstyle=\tiny\color{codegray},
    stringstyle=\color{codepurple},
    basicstyle=\fontsize{7}{7}\selectfont\ttfamily,
    breakatwhitespace=false,         
    breaklines=true,                 
    captionpos=b,                    
    keepspaces=true,                 
    numbers=left,                    
    numbersep=5pt,                  
    showspaces=false,                
    showstringspaces=false,
    showtabs=false,                
    tabsize=2
}
\lstset{style=listingstyle}

\section{Appendix}
\label{sec:appendix}

\subsection{Reflection Prompts}

Figures~\ref{fig:prompt_rag_repair} and~\ref{fig:prompt_non_rag_repair} show the prompts used for RAG-Repair and the non-RAG baseline respectively.

\begin{figure*}
\begin{lstlisting}[]
------------------------------------------------------------------------------------------------------------
RAG-Repair Prompt: 
------------------------------------------------------------------------------------------------------------

"""
You are an advanced reasoning agent that can improve based on self refection.
You will be given a previous reasoning trial in which you were given a question to answer.
You were unsuccessful in answering the question either because you guessed the wrong answer with Finish[<answer>] or there is a phrasing discrepancy with your provided answer and the answer key.
You will refer to the document {rag_man_result} for the failing command to find the solution for the issue.
You will refer to the document {rag_result} to find the solution for the issue.
The original Question presented by the user was {query}.
If you think you can solve the issue, first give me a kubectl command to run that helps me resolve this issue. Preface that command with "Command: " 
On the next line, clearly and concisely explain your reasoning in bulleted form. Use complete sentences.
You also need to consider that a failure can be due to typos in the Question presented by the user. In those cases, suggest possible corrections to the user based on the previous context and observation. Add it in your response. You may leverage kubeclt commands that allow you to list all namespaces first.
You used the following kubectlcommand: {cmd}
You got the following error: {error}

Begin! Remember that your answer must be in the form of a kubectl command.

Reflection: {agent_scratchpad}
"""
------------------------------------------------------------------------------------------------------------
\end{lstlisting}
\caption{Prompt for RAG-Repair}
\label{fig:prompt_rag_repair}
\end{figure*}

\begin{figure*}
\begin{lstlisting}[]
------------------------------------------------------------------------------------------------------------
Non-RAG-Repair Prompt: 
------------------------------------------------------------------------------------------------------------

"""
You are an advanced reasoning agent that can improve based on self refection.
You will be given a previous reasoning trial in which you were given a question to answer.
You were unsuccessful in answering the question either because you guessed the wrong answer with Finish[<answer>] or there is a phrasing discrepancy with your provided answer and the answer key.
The original Question presented by the user was {query}.
If you think you can solve the issue, first give me a kubectl command to run that helps me resolve this issue. Preface that command with "Command: " 
On the next line, clearly and concisely explain your reasoning in bulleted form. Use complete sentences.
You also need to consider that a failure can be due to typos in the Question presented by the user. In those cases, suggest possible corrections to the user based on the previous context and observation. Add it in your response. You may leverage kubeclt commands that allow you to list all namespaces first.
You used the following kubectlcommand: {cmd}
You got the following error: {error}

Begin! Remember that your answer must be in the form of a kubectl command.

Reflection: {agent_scratchpad}
"""
------------------------------------------------------------------------------------------------------------
\end{lstlisting}
\caption{Prompt for the non-RAG baseline}
\label{fig:prompt_non_rag_repair}
\end{figure*}

\subsection{Evaluation Dataset - Kubernetes Environment and Descriptives}

For our environment, we use a fresh initialization of minikube with the 0.33.8 version of the OpenTelemetry Demo (otel-demo)\footnote{https://opentelemetry.io/ecosystem/demo/} system installed. For our evaluation dataset, we only consider commands that are related to otel-demo because we need to be able to execute the commands in our evaluation Kubernetes environment. 

Guaranteeing the state of the Kubernetes environment for a large number of varied commands is difficult, so we choose to further filter the failures to exclude those that contain the phrase ``not found.'' This excludes error cases where particular resources are not found in the environment rather than issues with the command itself. For example, a query may request to get information about pod 042 which correctly generates the command `kubectl get pod 042' but the environment will still return an error due to the pod not being present in the environment. 

We use the baseline SRE-Agent used by ITBench~\cite{jha2025itbench} to generate the original \emph{kubectl} commands from user queries. In doing so, we also use a number of LLMs to generate \emph{kubectl} commands. In our final dataset, Llama3.1 8b instruct 128k generated 355 commands, GPT-4o generated 208, Llama3.3 70b 79, Granite3.2 8b 72, Granite3.1 8b instruct 128k 56, and GPT-o3 mini 2. We did not stratify for agent LLM when filtering the dataset, instead prioritizing error cases involving otel-demo. 

\subsection{LLMs Used and Size}

Table~\ref{tab:llmsused} lists the LLMs used by RAG-Repair and our non-RAG baseline in our experiments along with size in terms of active parameters.

\begin{table}[!htb]
\centering
\footnotesize
\begin{tabular}{@{}ll@{}}
\toprule
\textbf{Model}          & \textbf{Size in Active Parameters}                \\ \midrule
\textbf{Llama3.1 8b}    & 8B   \\
\textbf{Granite3.2 8b}  & 8B \\
\textbf{Mixtral 8x7b}   & 13B  \\
\textbf{Phi4}           & 14B    \\
\textbf{Llama4Maverick} & 17B \\
\textbf{DeepSeek-V3}    & 37B \\
\textbf{Mixtral 8x22b}  & 39B    \\
\textbf{Llama3.3 70b}   & 70B  \\
\textbf{Qwen2.5 72b}    & 72B   \\
\textbf{Mistral Large}  & 123B \\
\textbf{GPT-4o}         & Unknown  \\ \bottomrule
\end{tabular}
\caption{LLMs used in experiments in ascending order of model size with number of active parameters.}
\label{tab:llmsused}
\end{table}

\subsection{Correctness Categories}

We develop categories for our manual annotation to reflect how correctly the repaired \emph{kubectl} command addresses the given user query. We validated our categories by having two authors independently annotate the same 25 cases and calculated inter-rater reliability with a weighted Cohen's kappa of 0.79. 
Correctness Categories:
\begin{enumerate}
    \item \emph{Correct} - The user query was reasonably directly answered. As noted in the paper, there often are multiple correct methods of answering the same query.
    \item \emph{Partially Correct} - Only part of the user query was directly answered. This is particularly the case when a user query has multiple conditions and only one (usually the main one) is met. An example from our manual dataset is a query to forcefully drain nodes and target pods with a particular label. A partially correct repaired command was ``kubectl drain minikube --force --ignore-daemonsets --delete-emptydir-data'' which only does the first part of the query to forcefully drain nodes.
    \item \emph{Adjacent} - The result does not directly answer query but is a reasonable first step. In contrast to \emph{Incorrect}, this command can reasonably be part of a multi-step plan towards answering the query. An example from our manual dataset is a query to get information about a particular deployment ``opentelemetry-demo-grafana'' across all namespaces. Using the --all-namespaces parameter is not allowed by \emph{kubectl} in this case. An adjacent repaired command was ``kubectl get deployments --all-namespaces'' which just returns all deployments in all namespaces. A reasonable second step would be to filter those results for the given name.
    \item \emph{Incorrect} - The execution succeeded but did not answer the user query in any way. In contrast to \emph{Adjacent}, there is no reasonable way to use this command to answer the query. 
    \item \emph{Error} - The execution failed, specifically with a non-zero return code. We do not have any special consideration for failing commands that are almost correct but contain some minor error such as a hallucinated parameter.
\end{enumerate} 

\subsection{RAG Parameters}

By default, RAG-Repair uses the following parameters for RAG: a chunk size of 1,500, a distance metric of L\textsuperscript{2} (Euclidean distance squared), and the \emph{all-mpnet-base-v2} model for sentence embedding. The results as reported in subsection 4.3 use the default values listed. 

We test alternatives for each of these parameters across a subset of the tested models: Llama 3.1 8b, Llama3.3 70b, Llama4 Maverick, and Mixtral 8x22b. We test alternative chunk sizes of 500 and 800 which were found to be optimal sizes for chunking StackOverflow posts and documentation respectively in CodeRAG-Bench.\cite{wang-etal-2025-coderag}. We test an alternative distance metric of cosine similarity and two other sentence embedding models: Granite Embedding 107m Multilingual and Jina Embeddings v2. The latter embedding model is also trained on code. Lastly, we tested using the BM25 ranking function as an alternative to an embedding-based approach. The results are summarized in Table \ref{tab:ragresults}. We see that many of the alternative parameters seem to improve pass rate for Llama 3.1 8b but not other models. 

\begin{table*}[!bth]
\centering
\footnotesize
\begin{tabular}{@{}lllllll@{}}
\toprule
\textbf{Model} & \textbf{Chunk 500} & \textbf{Chunk 800} & \textbf{Cosine} & \textbf{BM25} & \textbf{Granite} & \textbf{Jina} \\
\midrule
\textbf{Llama3.1 8b}    & \textbf{0.67}           & \textbf{0.63}             & \textbf{0.63}   & 0.58 & \textbf{0.68} & 0.51       \\
\textbf{Llama3.3 70b}   & 0.51           & 0.53             & 0.55   & 0.51 & \textbf{6.3} & 0.49             \\
\textbf{Llama4Maverick} & 0.47           & 0.48             & \textbf{0.51}            & 0.40 & 0.44 & 0.40    \\
\textbf{Mixtral 8x22b}  & 0.44           & 0.42             & 0.45   & 0.38 & 0.39 & 0.28             \\
\bottomrule
\end{tabular}
\caption{Study of alternative RAG parameters, reporting \textit{pass rate} for six conditions: chunk size 500, chunk size 800, cosine similarity as a distance metric, using the BM25 ranking function, using the Granite 107m Multilingual embedding model, and using the Jina v2 embedding model. Bolded \textit{pass rate} are parameters that perform better than default RAG.}
\label{tab:ragresults}
\end{table*}

\end{document}